# Devil's staircase inside shrimp-shaped regions reveals periodicity of plateau spikes and bursts

Luiz F. B. Caixeta, Matheus H. P. Gonçalves, M. H. R. Tragtenberg, and Mauricio Girardi-Schappo[a)]
*Departamento de Física - Universidade Federal de Santa Catarina - Florianópolis SC - 88040-900 - Brazil*

Slow-fast dynamics are intrinsically related to complex phenomena and are responsible for many of the homeostatic dynamics that keep biological systems healthy functioning. We study a discrete-time membrane potential model that can generate a diverse set of spiking behavior depending on the choice of slow-fast time scales, from fast spiking to bursting, or plateau action potentials – also known as cardiac spikes, since they are characteristic in heart myocytes. The plateau of cardiac spikes can lose stability, generating early or delayed afterdepolarizations (EAD and DAD, respectively), both of which are related to cardiac arrhythmia. We show the periodicity changes along the transition from the healthy action potentials to these impaired oscillations. We show that while EADs are mainly periodic attractors, DADs usually come with chaos. EADs are found inside shrimp-shaped regions of the parameter space. However, in our system, multiple periodic attractors live within a shrimp-shaped region, giving it an internal structure made of infinite transitions between periodicities forming a complete devil's staircase. Understanding the periodicity of plateau attractors in slow-fast systems could be useful in unveiling the characteristics of heart myocyte behaviors that are linked to cardiac arrhythmias.

**Cardiac arrhythmias, a leading cause of heart failure, arise from disruptions in the timing and dynamics of cardiac myocyte action potentials, such as early and delayed afterdepolarizations (EADs and DADs). Understanding the periodicity changes in these action potentials (plateau spikes) is important to reveal the mechanisms behind these pathological conditions[1,2]. Using a discrete-time generic model, we reveal that these oscillatory phenomena are related to chaotic and periodic attractors. Some of the EAD attractors live inside shrimp-shaped regions in the parameter space, forming a complete devil's staircase of periodicity transitions before turning chaotic. This expands recent findings of quasiperiodic shrimps[3], although it contrasts with the original description of shrimp-shaped regions, which included only isoperiodic attractors[4,5]. Our insights could inform the study of membrane potential transitions during impaired bursting and plateau spiking, potentially enhancing diagnostics and guiding the development of therapies for cardiac dysfunction.**

## I. INTRODUCTION

Continuous-time conductance-based models pose a big challenge to theoretical studies because of the increased number of dynamical variables and free parameters. Simplified map-based models of the action potential (AP) can help unveil generic principles underlying the phenomenology of these complicated models. Thus, we study a simple and generic map-based AP model. Our discrete-time model for cardiac APs has three continuous state variables, six parameters, and a simple sigmoid transfer function. These features make its computational implementation trivial, efficient, and easily portable to any health and/or engineering application. The simplicity of this map-based model allows one to determine analytically most of the phase diagram. This model was recently used as a generic way to understand cardiopathologies[6], where different characteristics of the cardiac spike were linked to the underlying dynamics.

We describe the periodicity of the dynamics throughout the transition from plateau spikes to bursting, where the EAD-like and DAD-like oscillations linked to heart arrhythmias are found. We show the presence of shrimps that, instead of being isoperiodic, display an inner structure called a devil's staircase[7] along which the system transitions between infinitely many periodic solutions before reaching a chaotic attractor.

Although we physically interpret the model as the dynamics for single cell myocyte APs, it is worth noting that the equations were derived from an Ising model with competing interactions on a tree-like graph[8,9]. Our model can also be regarded as a mean-field approximations of either rate-based artificial neural networks[10,11], single neurons or dynamical perceptrons[12,13]. It was also used to study various nonlinear excitable phenomena either in isolation[14,15] or in coupled-map lattices, such as spiral waves[16,17].

In nonlinear dynamical systems, "*shrimps*" are originally a fractal distribution of regions of oscillatory attractors embedded within a chaotic sea on a bi-parameter space[4]. This offers rich insights into the stability and transitions between periodic solutions of a system[5]. Recently, this idea was extended to quasiperiodic shrimps[3], highlighting distinct dynamics such as torus-bubbling transitions and multitori attractors. Here, we unveil

[a)]Electronic mail: girardi.s@gmail.com

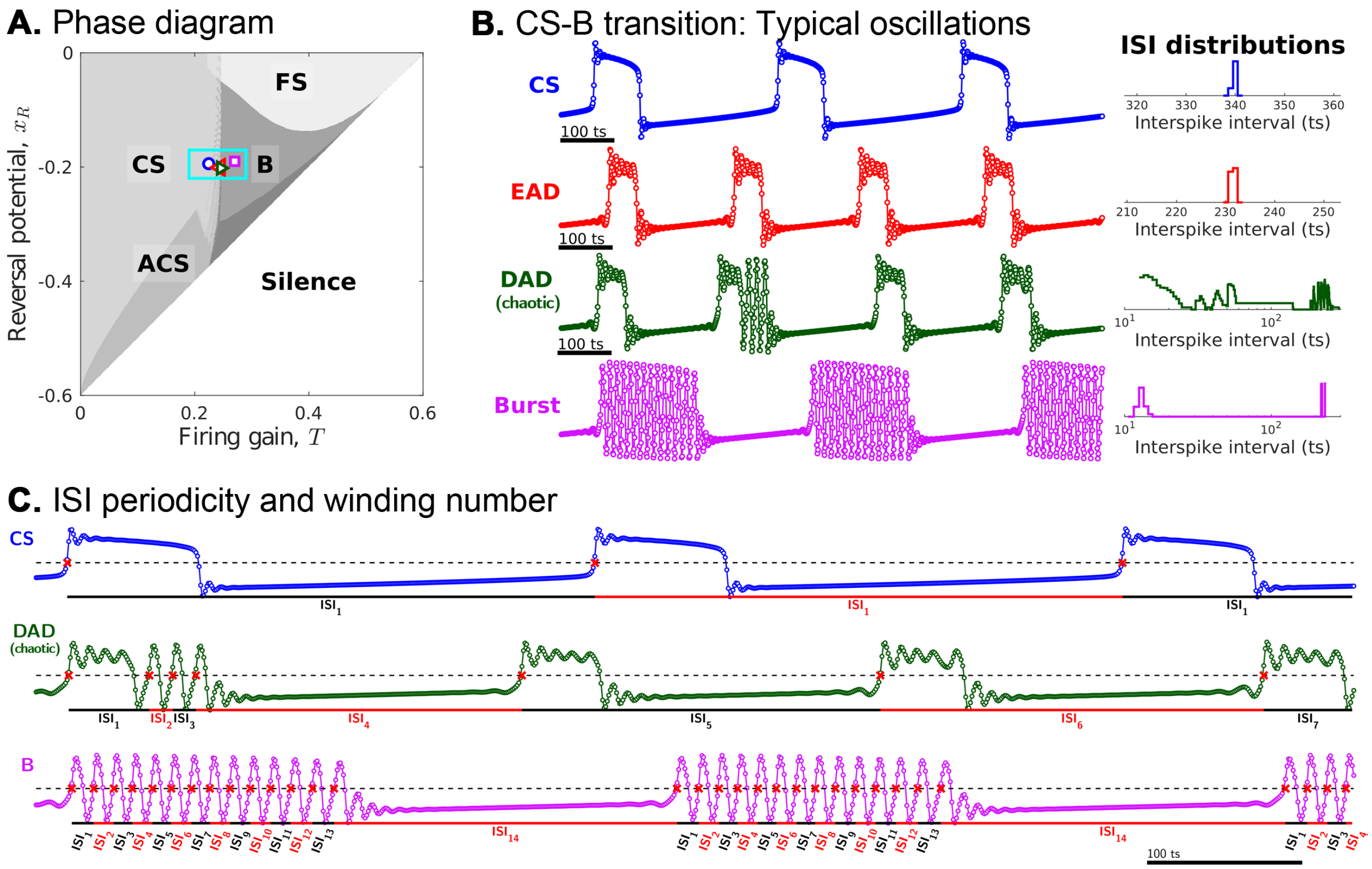

FIG. 1. **Phase diagram, oscillation modes and interspike interval. A.** Phase diagram coloring different ISI distribution profiles (see panel B, right) with shades of gray (see Methods). Notice a *dust*-like structure along the CS-B transition, which will be further investigated within the region highlighted by the cyan rectangle (detail shown in Fig. 3). From light to dark gray: fast spiking (FS), cardiac spiking (CS), aperiodic cardiac spiking (ACS), bursting spiking (B) as well as silence (hyperpolarized potential fixed points). Symbols: approximate selected values of $T$ and $x_R$ corresponding to the attractors shown in panel B: ○→CS; □→B; ▷→DAD; ◁→EAD. **B.** Example of oscillations found in the CS-B transition [$\circ \to x(t)$, left, solid lines are guides to the eyes only] with their typical ISI distributions (right). Parameters: $K = 0.6$, $\delta = \lambda = 0.001$ (fixed for all figures in this manuscript), and $T = 0.2248$, $x_R = -0.1942$ (CS); $T = 0.2447$, $x_R = -0.2005$ (EAD); $T = 0.2457$, $x_R = -0.2017$ (DAD); $T = 0.27$, $x_R = -0.19$ (B). **C.** Illustration of the definition of the sequence of ISI for each attractor. The period $P$ of the $\{ISI_n\}$ sequence reveals the number of cycles an attractor makes before repeating. Example of the attractors in panel B: CS → $P = 1$ because the same ISI is repeated successively; DAD → large $P$ limited by the total simulation time due to the aperiodicity of the chaotic attractor. B → $P = 14$, meaning that it takes 14 cycles for the repetition of the burst. Dashed line → $x = 0$ for reference; x → spike timestamp obtained using the conditions in Eq. (3).

shrimps that exhibit intricate internal structures in the form of stripes, with each stripe maintaining a periodic attractor. Striped structures in bi-parameter space are usual for systems having multiple stable periodic solutions[9,12,15,18], and are sometimes representative of Arnol'd tongues[9,12,15]. Remarkably, when analyzed along a single parameter, this collection of stripes forms a complete devil's staircase, providing a novel characterization of shrimp-related dynamics and further enriching the understanding of their organization within chaotic domains.

## II. MODEL

We study a discrete-time map with three variables. It was derived from an Ising model with competing interactions on a Bethe lattice[8,9]. Then, the hyperbolic tangent was simplified into a logistic function[15], $F(u) = u/(1 + |u|)$. Both this and the hyperbolic tangent are sigmoid functions. They increase monotonically with limits $F(u \to \pm\infty) = \pm 1$, and their first derivatives are continuous, where $F'(u) = 1/(1 + |u|)^2$. The advantages of this simplification are that all fixed points (FPs) become analytical and the computational cost to iterate the map is drastically reduced, preserving the rich repertoire of dynamical behaviors[15]. We interpret our model as the membrane voltage of a neuron[12,14] or a cardiac myocyte[6,13,15] over time. It is defined as

$$\begin{aligned} x(t+1) &= F\left(\frac{x(t) - Ky(t) + z(t) + H}{T}\right) \\ y(t+1) &= x(t) \\ z(t+1) &= (1-\delta)z(t) - \lambda(x(t) - x_R) \end{aligned}, \quad (1)$$

where $x(t)$ is the membrane potential of the cell at time $t$ with firing gain $T$; $y(t)$ is a fast feedback inhibitory potential coupled with conductance-like constant $K$ and $z(t)$ is the potential resulting from a slow current, such as the calcium dynamics[6]. The slow current has a recovery timescale $1/\delta$ and a driving timescale $1/\lambda$, with a reversal potential $x_R$. External inputs (synaptic or otherwise) can be introduced via the parameter $H$. All parameters and variables are given in arbitrary units.

This map has inversion symmetry, so that changing

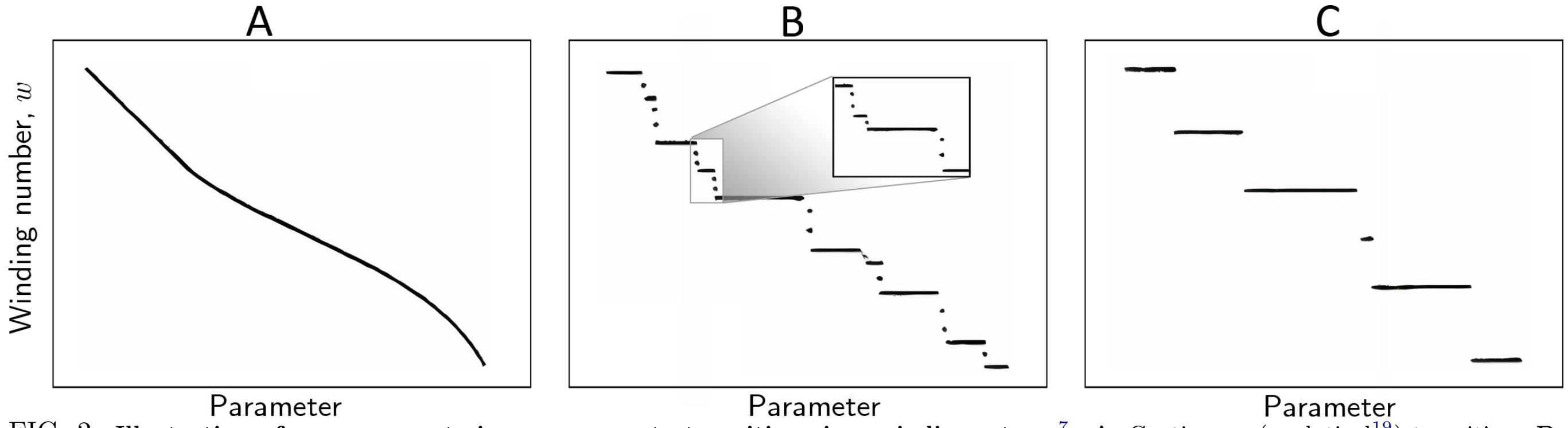

FIG. 2. **Illustration of commensurate-incommensurate transitions in periodic systems**[7]**.** **A.** Continuous (analytical[19]) transition. **B.** Complete devil's staircase. It has a (non-analytical[19]) fractal structure, such that the plateaus contain commensurate (periodic, $L < 0$) phases (rational $w$), and there are infinite plauteaus between every two plateaus. The incommensurate (quasiperiodic, $L = 0$) phases lie in between the commensurate phases (irrational $w$), making a set of zero measure in the parameter space. **C.** Harmless staircase. The system discontinuously transitions from one commensurate $w$ to another. The incomplete devil's staircase is similar to the harmless staircase, except that the transitions are continuous.

$x_R \to -x_R$ implies in changing $\mathbf{x}(t) \to -\mathbf{x}(t)$, where $\mathbf{x}(t) = [x(t), y(t), z(t)]$ is the solution of the map. Thus, we can choose $x_R < 0$ without loss of generality. We keep fixed the parameters $H = 0$, $K = 0.6$, and $\delta = \lambda = 0.001$, with initial conditions $x(0) = y(0) = z(0) = 1.0$. We use $x_R$ and $T$ as control parameters to trace the phase diagrams that delineate the oscillation modes of the map (Fig. 1A). Although the map produces a discrete set of points $x(t)$ for integer $t \geq 0$, we plot the attractors with interpolating lines to help visualizing the waveform of the oscillations (Fig. 1B,C).

The hyperbolic tangent model has a complete devil's staircase[8] as a function of $K$ with $T = 0.1$ and $H = \delta = \lambda = z(0) = 0$. The staircase becomes incomplete as $T$ grows[8,9]. Here, we are interested in characterizing the periodicity of the attractors along the transition from autonomous *cardiac spiking* (CS) to bursting (B) – see the $x_R \times T$ phase diagram in Fig. 1A. CS is also known as plateau spiking, as membrane depolarization lasts a very long time, forming a plateau[20,21]. This behavior is typical of heart myocytes[2]. Throughout the CS-B transition, the plateau loses stability via a delayed Neimark-Sacker bifurcation[6], generating either early afterdepolarization (EAD, Fig. 1B), or delayed afterdepolarization (DAD, unstable plateau followed by a quick burst of spikes in Fig. 1B). These forms of action potential are linked to cardiac arrhythmia[1,2].

A slow-fast analysis of our model can be performed in the limit[6] $\delta = \lambda \ll 1$. This is also known as adiabatic approximation. In this case, the variable $z(t)$ becomes slow when compared to $x(t)$, and so it can be turned into a parameter and absorbed inside the constant input $H' = H + z(t)$. This can be used to understand the emergence of cardiac oscillations in the model, since it can be shown that two stable fixed points coexist for[15] $H' = 0$, inside the region $T < 1 - K$ $(0 \leq K \leq 0.5)$ and $T < K - 2 + 1/K$ $(0.5 < K \leq 1)$. These fixed points give rise to a slow-fast hysteresis cycle as a function of $H'$, and the slow dynamics $z(t)$ makes the map go along this cycle[6].

## III. METHODS

The model undergoes an infinite-period bifurcation at[15] $x_R = -K + T$ (for small $\delta = \lambda$ and $H = 0$). Thus, we fixed $K = 0.6$, $\delta = \lambda = 0.001$, $H = 0$, and for each $(T, x_R)$ pair, we iterated Eq. (1) for $200,000$ time steps, discarding the initial transient of $20,000$ steps. Near the bifurcation, longer simulation times may be required to observe our results. We used $x(0) = y(0) = z(0) = 1.0$ as initial condition. Each attractor is characterized by three measurements: the sequence and distribution of the interspike interval (ISI), the maximum Lyapunov exponent $L$, and the associated winding number $w$ (*i.e.*, the ratio of cycles per period of the attractor).

### A. Interspike interval

The interspike interval is the number of time steps between two consecutive upswings of the membrane potential $x(t)$ (see Fig. 1C). More precisely,

$$ISI_n = t_{n+1} - t_n \tag{2}$$

where the instants $t_n$ and $t_{n+1}$ are defined by the simultaneous conditions

$$\begin{cases} x(t_k+1)x(t_k) \leq 0 & \text{[the map crossed } x=0 \\ & \text{between } t_k \text{ and } t_k+1] \;, \\ x(t_k) < x(t_k+1) & \text{[the oscillation is rising]} \;, \end{cases} \tag{3}$$

taking $k = n$ for the spike at time $t_n$ and $k = n+1$ for the spike at time $t_{n+1}$. There is no other $t$ between $t_n$ and $t_{n+1}$ that obeys both of these conditions. In other words, $t_n$ and $t_{n+1}$ can be regarded as the timestamps of consecutive spikes. Repeating this for every spike produces the sequence $\{ISI_n\}$ illustrated in Fig. 1C.

Note that a given attractor can have more than one unique ISI in $\{ISI_n\}$. This is the case for bursting, for example, where there are at least two distinct values in the sequence: the smallest corresponds to the interval between spikes within a burst, and the largest corresponds to the interval between spikes in consecutive bursts. This produces the multimodal distributions of $ISI_n$ shown for the Burst and DAD attractors in Fig. 1B–right. Thus, in general, an $ISI_n$ in the sequence is not the period of the attractor. Below, we explore the relation between the $\{ISI_n\}$ and the periodicity of the system.

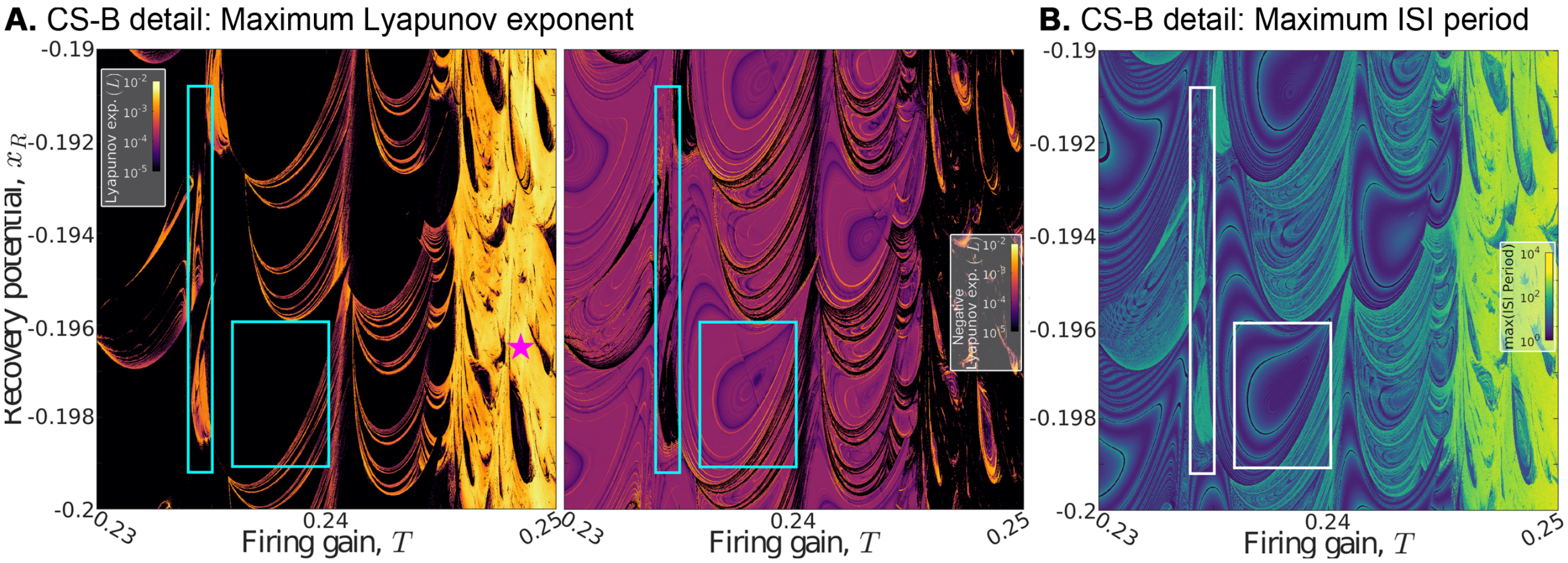

FIG. 3. **Detail of the *dust*-like structure in the CS-B transition. A.** Maximum Lyapunov exponent $L$. Colors → $|L|$ (positive $L$, left, and negative $L$, right); Color ranges: black-purple ($10^{-5} \le |L| < 10^{-4}$); purple-orange ($10^{-4} \le |L| < 10^{-3}$); orange-yellow ($10^{-3} \le |L| < 10^{-2}$). Left panel black gaps: non-chaotic regions (all the $L_i$ are negative). ⋆ → selected DAD attractor in Fig. 5B–bottom. **B.** The maximum period $P$ of the $\{ISI_n\}$ sequence closely matches the negative $L$ contour. Color ranges: blue shades ($1 \le P < 10$ ts); green shades ($10 \le P < 10^2$ ts); yellow shades ($10^2 \le P < 10^3$ ts). **All panels:** Rectangles → selected regions for further analysis in Figs. 4, 5 and 6. Taller rectangle → shrimp-shaped regions (Figs. 4C,E,F and 5); Smaller rectangle → Fig. 4B. Both rectangles also appear in Fig. 4A.

The discrete nature of time also introduces a ±1 variability for a given $ISI_n$ in the sequence $\{ISI_n\}$. This is because the conditions in Eq. (3) used to define $t_n$ and $t_{n+1}$ do not require the map to exactly repeat after the $t_{n+1}$ time step. Put differently, even if the waveform of the oscillation repeats after one ISI, the actual map value $x(t)$ does not need to do the same. This can be seen in the CS attractor in Fig. 1C: the attractor consists of the circles, and the spike timestamps $t_n$ are marked by a red "x" symbol which stands for $t_1$ in the first spike, $t_2$ in the second, and $t_3$ in the third. However, the values of the map [the circles marking $x(t_1+1)$, $x(t_2+1)$ and $x(t_3+1)$] following the timestamp are different for the three spikes (the distance of the circles to the $x=0$ dashed line increases during the spiking). This shifting of the map with respect to the waveform generates the ±1 variability that can be clearly seen in the ISI distribution for the CS and EAD attractors (Fig. 1B–right) – both of which have a steady waveform throughout which the actual values of the map slide.

We can illustrate this variability effect with the Poincaré section of a simple cosine function $v(t) = \cos(2\pi t/Q)$ with actual period $Q = 20$. Taking $v(t)$ only at integer $t \ge 1$, and applying it to the conditions in Eq. (3), produces the sequence $\{ISI_n\} = \{20, 19, 21, 19, 20, 21, \cdots\}$. By construction, we know the correct ISI should be equal to $Q = 20$, but values fluctuate. Sampling the series long enough, we can get $\langle ISI \rangle \approx Q = 20$ in this example. The equality $\langle ISI \rangle \approx Q$ holds only for periodic functions that repeat at every cycle.

If the attractor $x(t)$ is periodic, then the sequence $\{ISI_n\}$ must be periodic. This follows because if $x(t)$ is periodic of period $Q$, then $x(t) = x(t+Q)$ for all $t$. In particular, this is true for any $t = t_n$ in the sequence of upward crossings. In other words, if the map crosses $x = 0$ during a rise at $t = t_n$, it must also rise up at $t = t_{n+P} = t_n + Q$ after some integer number of cycles $P$. Therefore,

$$\begin{aligned} ISI_{n+P} &= t_{n+1+P} - t_{n+P} \\ &= t_{n+1} + Q - (t_n + Q) \\ &= t_{n+1} - t_n = ISI_n \ , \end{aligned}$$

and $ISI_{n+P+1} = ISI_{n+1}$, and so on and so forth, making $P$ the period of the $\{ISI_n\}$ sequence. Consequently, the period of the attractor is

$$Q = \sum_{k=n}^{n+P} ISI_k \ , \tag{4}$$

since $ISI_k$ is the duration of the $k$-th cycle of the map, and the map repeats after $P$ cycles. Also, if we iterate the map for a total of $mQ$ time steps ($m \gg 1$), the map executes $|\{ISI_n\}| = mP$ cycles, so the average ISI is

$$\langle ISI \rangle = \frac{1}{|\{ISI_n\}|} \sum_{k=1}^{|\{ISI_n\}|} ISI_k \approx \frac{mQ}{mP} = \frac{Q}{P} \ . \tag{5}$$

Since time is discrete in our model, we also plot the average ISI rounded to the nearest integer, $\lfloor \langle ISI \rangle \rceil$. The $\langle ISI \rangle$ quantity in Eq. (5) is only well-defined for periodic or quasiperiodic orbits. Chaotic attractors have no well-defined period $Q$ or number of cycles $P$, and hence the quantity $Q/P$ becomes arbitrarily dependent on the system details and simulation time.

We will show in the Results section that coloring the maximum period $P$ of the $\{ISI_n\}$ produces an internal structure in shrimps. This is seemingly contradictory with the original idea that shrimps are isoperiodic structures[4]. In an attempt to reconcile the shrimps in our model with the original idea, we show that the average ISI rounded to the nearest integer, $\lfloor\langle ISI\rangle\rceil$, has only two values inside each shrimp-shaped region.

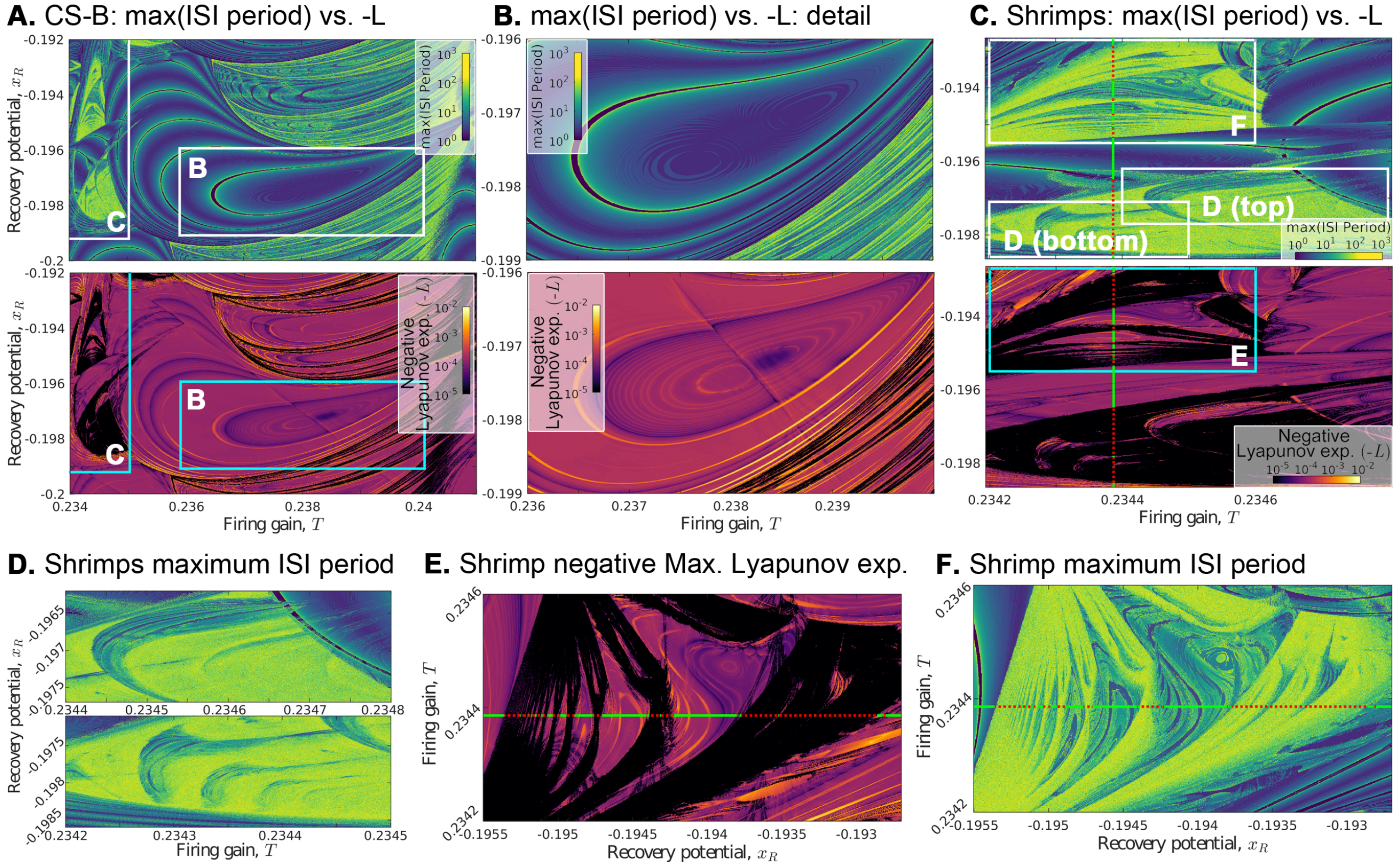

FIG. 4. **Shrimp-shaped regions internal structure.** Black to yellow colors → $-L$ (maximum negative Lyapunov exponent); ranges: black-purple ($10^{-5} \leq -L < 10^{-4}$); purple-orange ($10^{-4} \leq -L < 10^{-3}$); orange-yellow ($10^{-3} \leq -L < 10^{-2}$). Blue to yellow colors → $P$ (maximum ISI period); ranges: blue shades ($1 \leq P < 10$ ts); green shades ($10 \leq P < 10^2$ ts); yellow shades ($10^2 \leq P < 10^3$ ts). A very similar contouring is obtained by both metrics (as also seen in Fig. 3A–right) suggesting a relation, see Section IV E. **A.** Detail of Fig. 3B. The non-chaotic region shows rings of constant $P$. Rectangles → regions displayed in panels B and C. **B.** Detail of the non-chaotic region forming rings of constant $P$ (smaller rectangle of panel A and Fig. 3). **C.** Region with shrimps (tall rectangle of panel A and Fig. 3). Rectangles → regions displayed in panels D, E and F. **D, E, F.** Periodicity changes inside shrimps forming stripes of constant $P$. **E, F.** Selected shrimp sequence plotted as $T \times x_R$ (see also Fig. 5C). Horizontal line: selected $T = 0.2343864$ for the bifurcation analysis vs. $x_R$ (bifurcation diagrams of Figs. 6, 7 and 8); green solid line → non-chaotic regions inside shrimps; red dotted line → chaotic regions in between shrimps. Shrimp stripes → periodicity steps in the devil's staircase (see Section IV D).

### B. Winding number

Generally, attractors can be periodic in a way that $Q$ and $P$ share no common factors. And this is well-understood in terms of the winding number $w = P/Q$ (number of cycles $P$ per period $Q$). The winding number $w$ for periodic orbits is defined as the number of cycles $P$ executed by the map during a single period of oscillation $Q$. The derivation of Eq. (5) implies that

$$w = \frac{P}{Q} = \frac{1}{\langle ISI\rangle} \ . \tag{6}$$

Rational $w$ implies in periodic phases where $P$ and $Q$ are commensurable, *i.e.*, we can always find an interval of time $mQ$ inside which lie $mP$ cycles of the oscillation. When $P$ and $Q$ share no common factors, the periodic oscillation can be said to exist in a torus-shaped phase space[22]. Thus, analyzing the non-rounded $\langle ISI\rangle$ must be equivalent to analyzing $w$. Conversely, irrational $w$ implies in an incommensurate (quasiperiodic) oscillation in the torus[22]. Eq. (6) can be extended for this case[7,22–24]. Incommensurate phases can be sliding or locked in the $(-1; 1)$ ranged mapped by the sigmoid $F(u)$. The sliding case means that $x(t)$ can take any value within a connected subinterval of this range. A locked attractor exist only in a disconnected subinterval of this range.

The transition between commensurate and incommensurate phases has been studied in the context of spatial ordering in magnets and other systems[7,19,24–26], including in the original version of our model where $F(u) = \tanh(u)$ and $x(t)$ is the magnetization at inner layer $t$ of the Bethe lattice[8,9]. As a single parameter is varied, the transition can be continuous, discontinuous, or quasi-continuous (see Fig. 2). Here, we will study $w$ as a function of $x_R$ with all the other parameters kept fixed.

A continuous transition is known as *analytical*[19], where the system goes smoothly from one $w$ to another without staying trapped in a single $w$ as the parameter is varied (Fig. 2A). On the other end, discontinuous transitions are characterized by the system jumping from a commensurate phase to another, and there are only finitely many commensurate phases in the considered parameter range (Fig. 2C). A complete devil's staircase is when there are infinitely many commensurate phases in the considered parameter range, and hence $w$ varies non-analytically (Fig. 2B). The steps in a complete devil's staircase can make a nonstandard Farey sequence[12,27]. This means that, given two steps, one with $w_1 = P_1/Q_1$ and the other with $w_2 = P_2/Q_2$, there is a third step in between, with

$$w_3 = \frac{mP_3}{mQ_3} = \frac{P_3}{Q_3} \ , \tag{7}$$

such that $mP_3 = P_1 + P_2$ and $mQ_3 = Q_1 + Q_2$, and $m$ is a positive integer common factor. The sequence is *nonstandard* for some $m > 1$. This construction in Eq. (7) holds for any two steps in the staircase. Note that $m$ does not need to be constant during the construction of the $w_k$ sequence. In particular, for $m = 2$, this means that there are two coexisting attractors, one has solution $\mathbf{x}(t) = [x(t), y(t), z(t)]$, and the other has solution $-\mathbf{x}(t)$. The steps make a dense set in the parameter space.

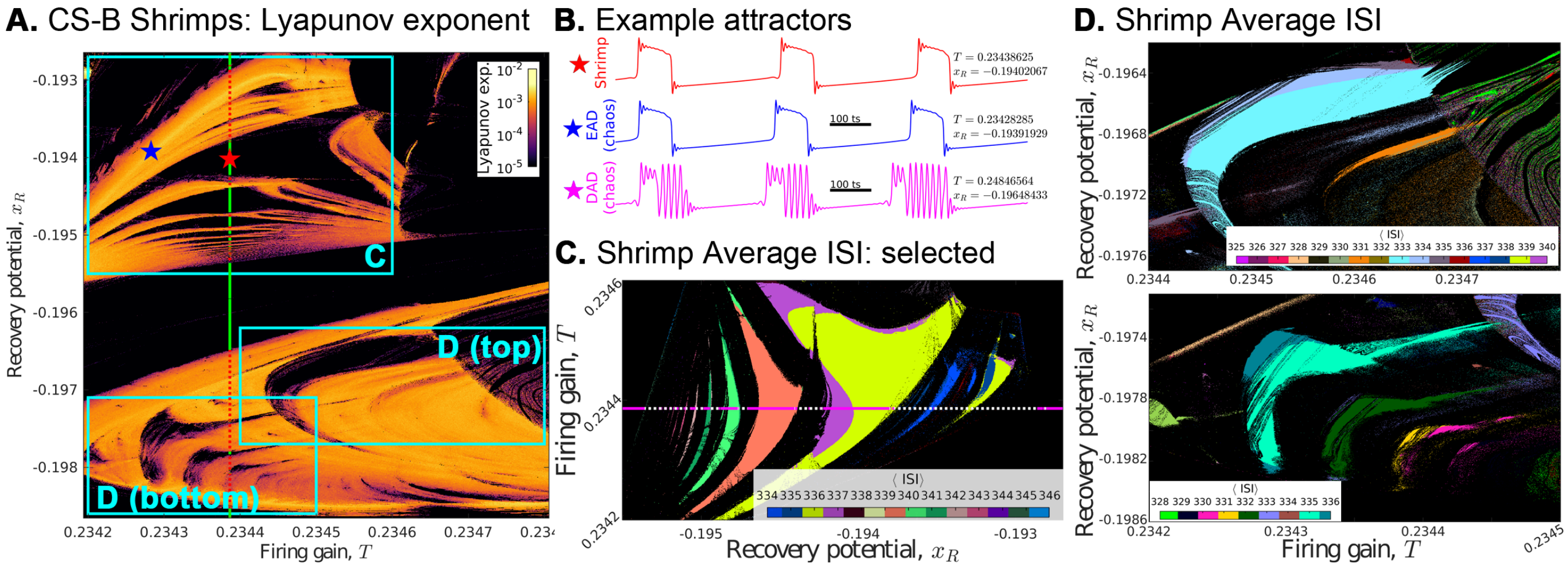

FIG. 5. **Shrimps viewed with rounded average ISI. A.** Shrimps are non-chaotic regions (black → $L \leq 0$; detail of taller rectangle in Fig. 3). Rectangles → regions in panels C and D. ★, ★: selected attractors for panel B. **B.** EADs show up inside (top) and outside (middle) shrimps. DADs appears in chaotic regions (bottom, ★ in Fig. 3A). **C,D.** $\lfloor\langle ISI\rangle\rceil$ ($\langle ISI\rangle$ rounded to the nearest integer – see Methods) shows a single value in the bulk of shrimp-shaped regions; the borders show a secondary $\lfloor\langle ISI\rangle\rceil$. **C.** Selected sequence of shrimps plotted as $T \times x_R$ for further inspection (Figs. 4 and 6). Horizontal line: selected $T = 0.2343864$ for the bifurcation analysis vs. $x_R$ (bifurcation diagrams of Figs. 6, 7 and 8); ━ → non-chaotic regions inside shrimps; ⋯⋯ → chaotic regions in between shrimps.

When there is a complete devil's staircase, incommensurate quasiperiodic phases lie in a set of zero measure in the parameter space that is complementary to the set generated by the steps. Thus, the complete devil's staircase is a sort of Cantor fractal[24]. We estimate the fractal dimension $D_f$ by the standard box-size scaling procedure[28]: the plot is divided into boxes of size $s$ and we count the number of boxes that contain a data point, and repeat this for various box sizes $s$. The slope of the log-log count vs. $s$ curve is approximately $D_f$ (see Appendix B for details).

We will show that, inside shrimps, $w$ makes a complete devil's staircase, in which the system jumps between infinitely many periodic orbits before going chaotic. This expands recent results in a predator-prey system where quasiperiodic orbits were found inside shrimps[3]. This also suggests that shrimps are more general than originally thought, and can contain: (a) isoperiodic orbits[4,5]; (b) infinitely many periodic orbits (our study); or (c) quasiperiodic orbits[3]. Isoperiodic shrimps are found only in regions where $P$ and $Q$ stay constant as the parameters are varied.

### C. Lyapunov exponents

The Lyapunov exponents measure the rate of phase space volume contraction/expansion of the system over time. There is one exponent $L_i$ for each phase space direction. If the largest exponent, $L = \max_i L_i$ is greater than zero, $L > 0$, the attractor is said to be chaotic. On the other hand, the case where $L \leq 0$ can occur for both periodic and quasiperiodic orbits. A maximum Lyapunov exponent $L = 0$ happens for quasiperiodic orbits and at the transition boundary to chaos. The case $L < 0$ corresponds to stable dissipative systems and/or attractive periodic orbits. We calculate $L_i$ using the Eckmann-Ruelle method[29], Eq. (A3), which we derive in Appendix A.

## IV. RESULTS

The phase diagram of oscillation modes, Fig. 1A, was colored by classifying the attractors through their ISI distribution (see[15] for details). This revealed a *dust*-like structure in the boundary between the CS and B transition. CS transforms into B passing through a delayed Neimark-Sacker bifurcation, making the plateau unstable and, eventually, generating afterdepolarization spikes[6]. These transition attractors, labeled as EAD and DAD in

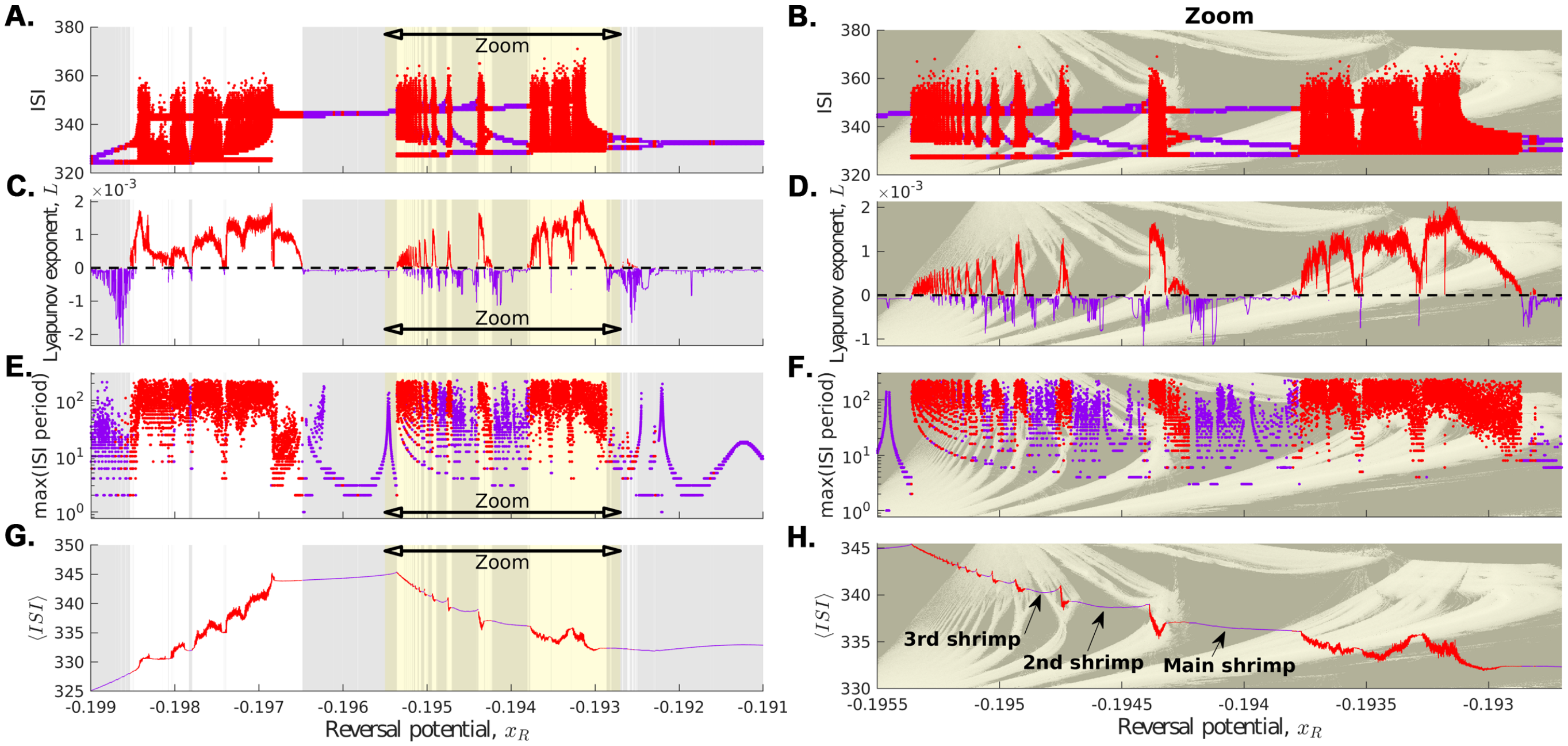

FIG. 6. **Shrimps' ISI diagrams for fixed** $T = 0.2343864$**.** Left column→traversing the parameter space over the solid-dotted lines shown in Figs. 4C and 5A (darker background → non-chaotic regions where all Lyapunov exponents are negative). Left column yellow background → selected $x_R$ range shown on the right column. Right column → detail of the left column, traversing the shrimp sequence shown in Figs. 4E,F and 5C along the horizontal dotted-solid line shown there. We plot the shrimps' silhouette on the background highlight which of the shrimps is being crossed. Red curves→positive maximum Luapunov exponents (chaotic attractors); purple curves→all Lyapunov exponents are negative. **A and B.** ISI bifurcation diagrams showing that attractors inside the shrimps (darker yellow background) can have more than one value in the $\{ISI_n\}$ sequence (*e.g.*, the main shrimp has three distinct ISI's between 330 and 350 ts). **C and D.** The maximum Lyapunov exponent shows chaotic regions in between shrimps. **E and F.** The $\{ISI_n\}$ has multiple periods $P$ inside shrimps (ranging from $P = 1$ where attractors repeat at every oscillation, to $P \sim 200$). **G and H.** $\langle ISI \rangle$ is shown without rounding, revealing a slight variation inside shrimps (pointed by arrows, further analyzed in Fig 7), and forming a harmless staircase going from one shrimp to another.

Fig. 1B,C, are present in heart myocytes when patients display cardiac arrhythmias[1,2]. In our map, the DADs that we observed are chaotic, whereas EADs may be periodic or chaotic depending on the parameters as we will show. Here, we investigate the structure of the dusty region in the transition between CS and B, and characterize the periodicity of its corresponding attractors.

### A. ISI sequence reveals structure of non-chaotic regions

Zooming in the dust using the Lyapunov exponent, we can see islands of non-chaotic behavior forming twisted half-moon shapes (Fig. 3A–left). These shapes prevail for $T < 0.243$, which is the value of $T$ in which oscillations first appear in the slow-fast description of the model[6] obtained when $\delta = \lambda \ll 1$. The maximum period of the $\{ISI_n\}$ reveals that these non-chaotic regions form rings of isoperiodic ISI sequences (Fig. 3B). These rings closely match the ring patterns in the maximum Lyapunov exponent $L$ when $L < 0$ (Fig. 3A–right). Note, however, that the regions with $L > 0$ are chaotic. We also plot the period $P$ of the $\{ISI_n\}$ of these chaotic attractors for a better visualization of the diagram, even though we acknowledge that $P$ is undefined in this case. Within chaotic regions, $P$ appears large, but $P$ is actually dependent on the simulation time. The more time we iterate the model, the larger the period of the ISI sequence of chaotic attractors will be due to their aperiodic nature.

Squeezed between the half-moon shapes, there are different sequences of shrimp-shaped regions (Figs. 4 and 5). Originally, shrimps were proposed as fractal regions in the parameter space where attractors have the same period[4,5]. It is worth noticing that we refer to these structures as *shrimps* because they resemble those originally found in the Hénon map[4]. However, our map has three variables, a slow-fast dynamic and a cubic nonlinearity due to the sigmoid shape of $F(u)$, whereas the Hénon map is a two-variable quadratic equation with a single time scale. Also, differently from the Henón map, the oscillatory behavior of the attractors we are studying (CS, EAD, DAD and B) intrinsically require the slow-fast dynamic to exist. Hence, our attractors can have multiple ISI in the $\{ISI_n\}$ sequence, potentially leading to a distinct phenomenology. We selected three sequences of shrimps to further explore in details throughout the rest of the manuscript.

### B. Shrimps have multiple periodicity of the ISI sequence

The non-chaotic half-moon shapes have internal rings of constant $P$, and this pattern is closely followed by the Lyapunov exponent $-L$ (Figs. 4B). This suggests that the non-chaotic regions inside shrimps might also have some internal structure. We looked inside the shrimp-shaped regions using both the maximum period $P$ of the $\{ISI_n\}$, and the maximum Lyapnov exponent $L$ when all $L_i$ are negative (Fig. 4C,D,E,F). Similarly to what happens inside the half-moon regions, shrimps show multiple periods of the $\{ISI_n\}$ sequence (Fig. 4C,D,E,F). However, the isoperiodic regions of $\{ISI_n\}$ form stripes inside the shrimps instead of rings. These stripes correspond to steps in the devil's staircase (discussed in Section IV D). Inside the shrimps, a similar striped pattern appears in the maximum Lyapunov exponent $L$ when it is negative, $L < 0$ (Fig. 4–all panels). This suggests that $P$ and $L$ are related (see Section IV E).

### C. A single rounded average ISI prevails inside shrimps

Fig. 5A shows that shrimps are not chaotic, since they have $L \leq 0$, and are surrounded by chaotic attractors (shown in colors). In the original definition, the whole region within a shrimp in the Hénon map contains attractors with the same period[4]. However, we observed that, in our model, this is not true. Since time is discrete and attempting to reconcile our findings with the phenomenology of the Hénon map[4], we plotted the shrimp-shaped regions of the phase diagrams coloring the average interspike interval rounded to the nearest integer, $\lfloor\langle ISI\rangle\rceil$ (Fig. 5).

The rounded average ISI of the attractors results in a single value $\lfloor\langle ISI\rangle\rceil$ in the bulk region of each shrimp structure [Fig. 5C,D]. The boundaries of the shrimp-shaped regions have a secondary rounded average ISI, $\lfloor\langle ISI\rangle\rceil$. Zooming in on these regions in Fig. 5C,D, we observe that the shrimp structures decrease in size while following a sequence of values $\lfloor\langle ISI\rangle\rceil$ that slowly increase for each progressively smaller shrimp. This zoomed-in analysis highlights a fractal-like hierarchy in the shrimp structures, revealing an intricate temporal organization of ISI as the shrimps scale down. However, this does not correspond to the typical period-doubling cascade found in isoperiodic shrimps[4,5].

### D. ISI bifurcation and the devil's staircase

In order to detail the dynamics inside and around the shrimps, we fixed $T = 0.2343864$ and varied $x_R$ throughout the main shrimp sequence shown in Figs. 4C,E,F and 5A,C. Along this fixed $T$ line, we plotted the ISI bifurcation diagram (Fig. 6A,B), the maximum Lyapunov exponent (Fig. 6C,D), the maximum period $P$ of the ISI sequence (Fig. 6E,F), and the average ISI (Fig. 6G,H).

All the shrimps in the sequence show three or more distinct ISIs (Fig. 6A,B) ranging from, approximately, 330 to 350 time steps. Some $ISI_n$ in the sequence starts branching out inside shrimp regions as the in-between chaotic attractors are approached (maximum $L$ shown in Fig. 6C,D). This can be clearly seen for the three largest shrimps: in the smaller ones, the upper ISI starts branching out as $|x_R|$ decreases towards the largest shrimp on the right (Fig. 6B). This organization is intimately related to the nearby chaos. This is because the periodic regions farther from the shrimps (darker shaded regions in Fig. 6A) have single ISIs.

The period of the $\{ISI_n\}$ sequence is shown in Figs. 6E,F. As explained in Methods, each $ISI_n$ in the sequence may show a slight variation of $\pm 1$ due to the discrete nature of the map with respect to the waveform. This means that some of the data shown in these panels are due to these random fluctuations. The maximum period of the sequence is more reliable because the period due to random fluctuations become negligible as the period of the sequence grows. Farther from the shrimp region, the periodic regions display a more well-behaved period of the ISI sequence. Inside the shrimps, the period $P$ of the ISI sequence shows multiple values, although not as many as in the chaotic regions (since $P$ is ill-defined).

We show the $\langle ISI\rangle$ without rounding in Figs. 6G,H. The regions in between shrimps show a spurious $\langle ISI\rangle$ due to the aperiodic behavior of the chaotic attractors. Let us analyze in detail the larger shrimps, starting from the largest one (main), and going to the left (the direction of decreasing $x_R$). We can see that the $\langle ISI\rangle$ takes discrete steps from one shrimp to the other. Rounding to nearest integer, $\lfloor\langle ISI\rangle\rceil$, this becomes analogou to steps in a harmless devil's staircase, explaining the ISI increasing sequence shown in Fig. 5C. However, checking the raw values of the $\langle ISI\rangle$, there is a subtle slope inside the shrimp. This is particularly noticeable in the first three shrimps pointed by arrows (Fig. 6H).

Fig. 7 shows the detail of the $\langle ISI\rangle$ for the first three shrimps compared to the inverse winding number $1/w = Q/P$. The whole extent of each panel is inside a shrimp, so the staircases appear as features of the shrimps themselves. We expect that $\langle ISI\rangle = 1/w$ [Ex. (6)]. In the three panels, we can se that both the $\langle ISI\rangle$ and the $1/w$ curves match almost completely. However, as discussed in Methods, the intrinsic fluctuations in each $ISI_n$ have to be counteracted. Thus, we obtained a better match between both quantities by plotting $\langle ISI\rangle - 1$ for comparison.

The $\langle ISI\rangle$ is much easier to measure, and provides a great estimate of $1/w$, even though the precise fractal dimensions of the staircase and the attractor labels for each step can only be obtained from the $w$ data. We found fractal dimensions of about $D_f \sim 0.95$ for the main shrimp and $D_f \sim 0.98$ for the other two, meaning that the staircase is tightly packed with periodic attractors, almost resulting in a continuous line. The gaps in the staircases are due only to the finite simulation time (notice that $Q$ is very large).

The whole staircase fits inside one unit of $\langle ISI\rangle$ because the period $Q$ of each attractor is relatively large compared to the number of cycles $P$. This is a direct consequence of the slow time scale that makes up the plateau spikes, $\delta = \lambda = 10^{-3}$ ts$^{-1}$. Increasing $\delta = \lambda$ destroys the plateaus and would possibly disrupt the findings that we are describing here. We highlighted a few winding number $w$ labels in each shrimp to show that the staircase generates a nonstandard Farey tree sequence. For example, the three labeled steps in the main shrimp have $w_1 = 6/2016$, $w_2 = 3/1007$ and $w_3 = (6{+}3)/(2016{+}1007) = 9/3023$ (the step between $w_1$ and $w_2$). A similar pattern happens for the three labeled steps in the second shrimp. In the third one, however, we see that $w_1 = 25/8486$ (left step), $w_2 = 140/47524$ (right step), yielding

$$w_3 = \frac{25+140}{8486+47524} = \frac{15\times 11}{15\times 3734} = \frac{11}{3734}\,,$$

showing the nonstandard nature of the sequence, since $m = 15 > 1$ for this particular step. The $w_1$, $w_2$, $w_3$ relation can be applied to any three steps in the shrimps, provided that $w_3$ is a step between $w_1$ and $w_2$. The second and third shrimps (Fig. 7B,C) have non-monotonic staircases, something that we have not seen in any other model.

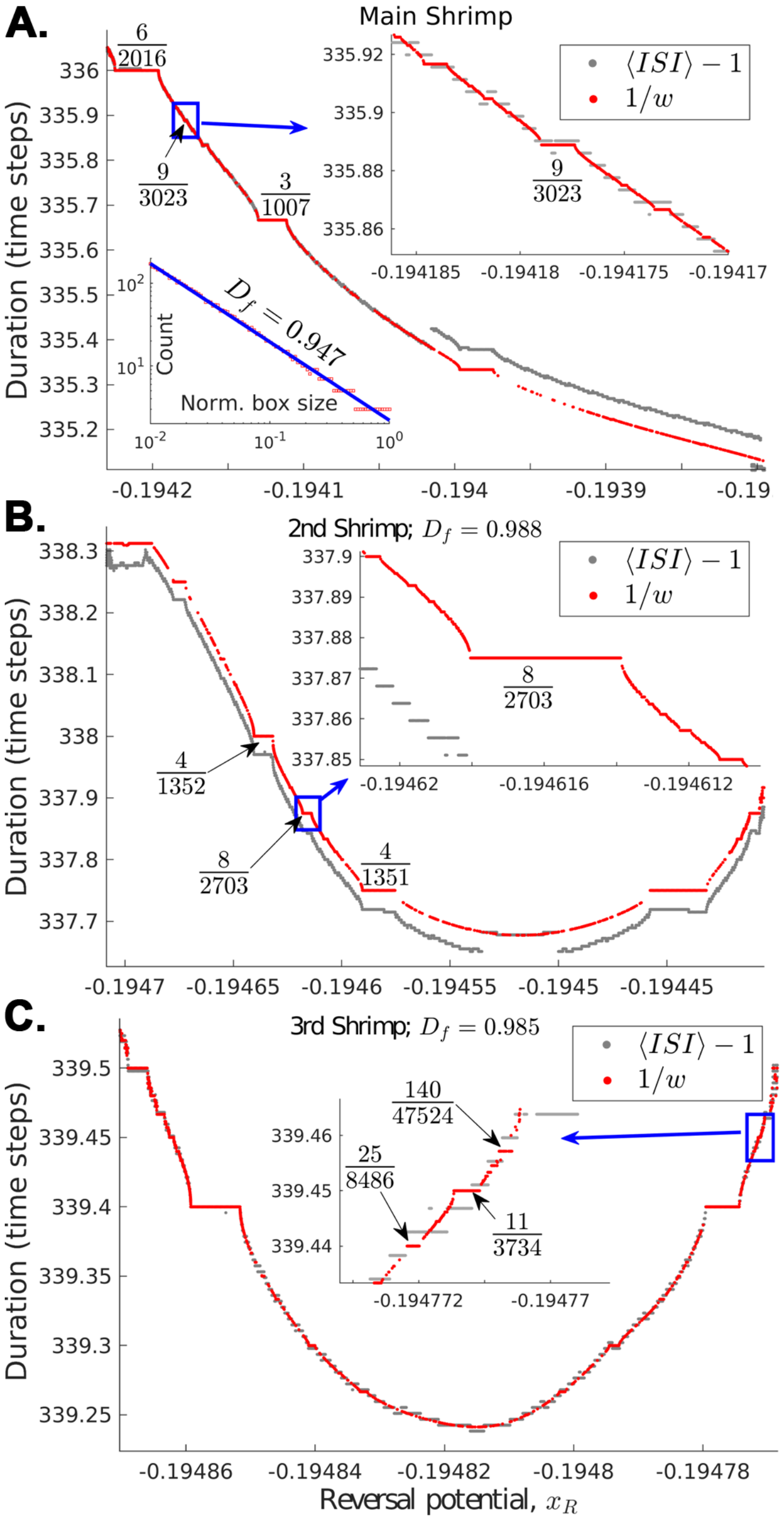

FIG. 7. **Shrimps' complete devil's staircases.** Detail of the $\langle ISI \rangle$ data pointed by arrows in Fig. 6H compared to the inverse winding number, $1/w = Q/P$. Some staircase steps are labeled by $w = P/Q$ to highlight the Farey tree structure of the system. Each panel has $x_R$ crossing the shrimp at fixed $T = 0.2343864$. $\langle ISI \rangle - 1$ is approximately equal to $1/w$ due to the $\pm 1$ intrinsic variability of the $\langle ISI \rangle$ discussed in Methods. The steps in the staircase are the stripes of the maximum ISI period and maximum negative Lyapunov exponent shown in and 4E,F. The analysis here is valid for all the shrimps that we show in this manuscript. **A.** Main shrimp (largest one in the sequence). **A (bottom inset).** Box count metric used to estimate the fractal dimension of the staircase. **B.** Second shrimp (first to the left of the main one). **C.** Third shrimp (second to the left of the main one). **Top insets.** Detail of the main panel inside the drawn rectangle.

### E. Relation between the Lyapunov exponent and the number of cycles near quasiperiodic orbits

Quasiperiodic attractors are associated with maximum Lyapunov exponent $L = 0$. Meanwhile, the maximum ISI period $P$ has to diverge for these orbits, since they never repeat. This happens concomitantly with the divergence of $Q$, leaving an irrational $w$ at the boundary between periodic phases. Each step on the devil's staircase comprises periodic orbits in which both $L < 0$ and $P$ is constant and finite. As we walk the staircase approaching the boundary of a step, $-L$ decreases towards zero and $P$ only diverges at the boundary of the steps due to the non-analytical nature of the complete devil's staircase. This happens systematically, such that at boundary between steps, both $L = 0$ and $P \to \infty$, since the periodic orbit gives place to a quasiperiodic attractor. This suggests that $-L$ and $P$ are related by

$$-L \sim \frac{1}{P} \tag{8}$$

near quasiperiodic orbits inside regions of the parameter space where the system displays a complete devil's staircase. Note that we are not claiming that both $P$ and $L$ are analytically dependent on one another, so we did not use an equal sign in Eq. (8). Instead, both quantities can behave inversely proportional to each other, generating a correlation between $P$ and $L$ along the staircase.

In Fig. 8, we plot $-L$ and $1/P$ along the $x_R$ axis showcasing parts of the devil's staircases in each of the three shrimps that we are analyzing. It is easy to see that the maxima of both of the curves coincide throughout the staircase: notice that the plateaus of $1/P$ (steps on the staircase) coincide with the peaks of $-L$. In fact, the correlation coefficient between these two curves varies from $R = 0.74$ to $R = 0.91$, depending on the chosen shrimp (see Appendix C for details). All correlations are significant ($p = 10^{-6}$ in simple t-statistic estimate from $R$). Moreover, both the units of $-L$ and $1/P$ are related to inverse time units. These findings give strong support to the relation in Eq. (8), even though we did not derive it from first principles.

Every step (periodic region, finite non-zero $-L$ and $1/P$) is enclosed between two quasiperiodic points in pa-rameter space ($L = 0$, $P \to \infty$) along a single parameter axis, $x_R$, for fixed $T$. Varying $T$ a little generates a slightly different set of $x_R$ points where the quasiperiodic orbits are found (again, defined by both a zero $L$ and diverging $P$). If we continue this process for every $T$ in the bifurcation diagrams, the striped pattern shown in Figs. 3 and 4 emerges for both $-L$ and $P$ (the maximum $\{ISI_n\}$ period). This is because each quasiperiodic point in the single parameter space generates generates a contour line in bi-parameter space. The enclosed periodic region in the single parameter space (*i.e.*, the devil's staircase step), on the other hand, gives rise to a stripe in the bi-parameter space, explaining the similar heatmaps found both for $-L$ and $P$.

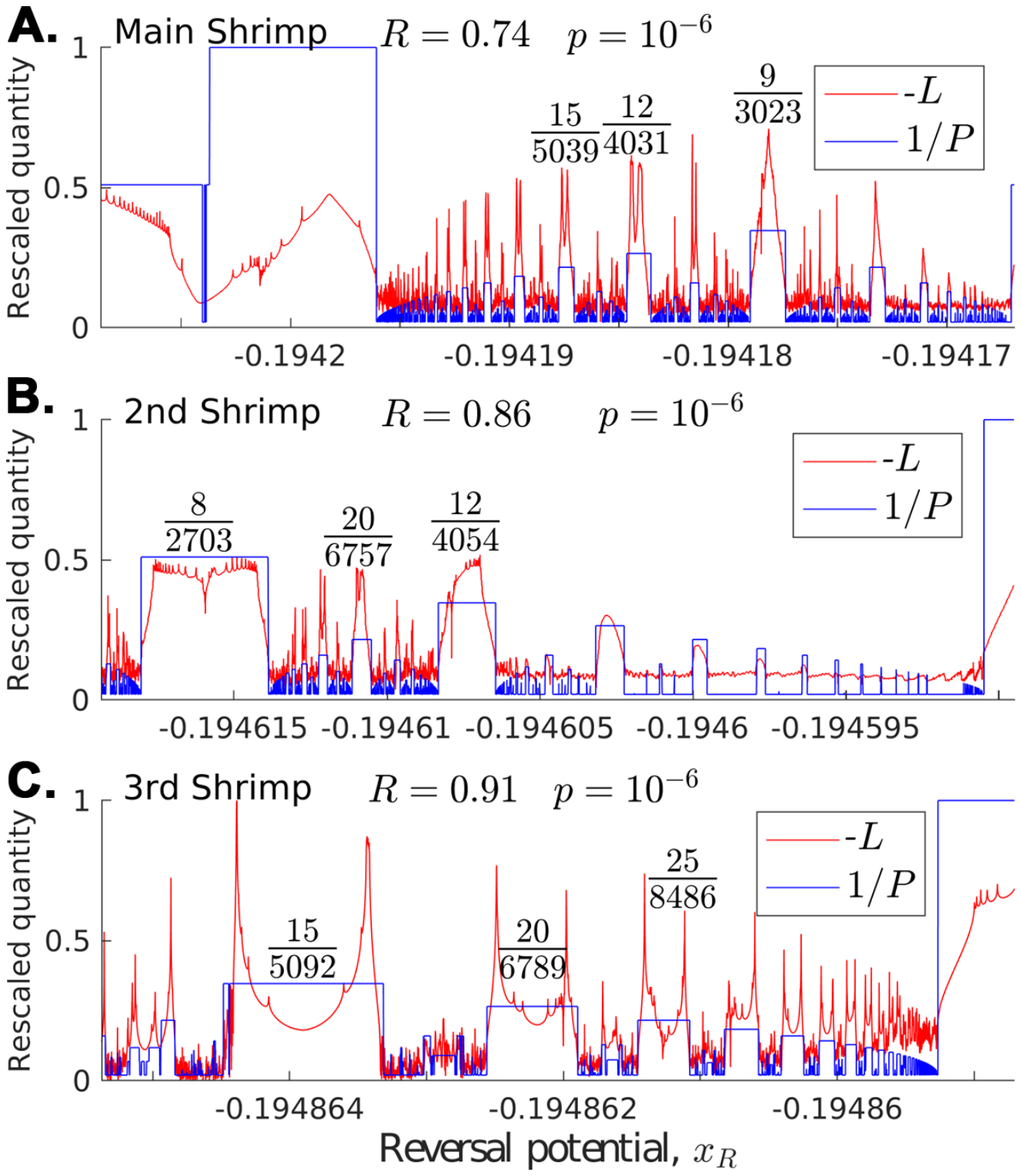

FIG. 8. **Relation between the maximum negative Lyapunov exponent and the maximum ISI period.** Each panel is a detail of each devil's staircase presented in Fig. 7 (also pointed by arrows in Fig. 6H). We cross each shrimp by varying $x_R$ at fixed $T = 0.2343864$. (red line) → maximum Lyapunov $L$ (since it is negative, we plot $-L$). (blue line) → inverse of the maximum period $P$ of the $\{ISI_n\}$ sequence (diverging $P$ was replaced by $10^6$, an arbitrary constant that is much larger than every other $P$, for numerical reasons). Each plateau in $1/P$ corresponds to a step in the staircase, strongly correlating with maxima in $-L$. Some plateaus are labeled by $w = P/Q$ to emphasize the nonstandard Farey sequence of periodicities between steps. **A.** Main shrimp. The correlation coefficient between $-L$ and $1/P$ is $R = 0.74$. **B.** Second shrimp, $R = 0.86$. **C.** Third shrimp, $R = 0.91$. All correlations are significant (p-statistics: $p = 10^{-6}$ for a simple regression of $-L$ vs. $1/P$, see Appendix C). Only for plotting, both quantities have been rescaled to the [0; 1] range.

## V. CONCLUSION

We studied a three-variable map that can be employed in different areas: from magnets to membrane voltage models. In particular, we interpret $x(t)$ as a membrane potential, such that the homeostatic field $z(t)$ introduces a slow-fast dynamic that is capable of generating plateau spikes and bursts. The transition between these regimes is permeated by a loss of stability of the plateau, generating early and delayed afterdepolarizations of the membrane. This behavior is found in some cardiac arrhythmias due to impairment in ionic channels[30,31]. For example, delayed sodium currents can prolong the AP, enabling calcium currents to destabilize repolarization and cause EADs[32,33]. Sodium-triggered EADs and DADs can occur without altering AP duration[34]. Compromised slow potassium currents are critical for AP prolongation and the emergence of EADs or DADs[30,31].

In our model, slow currents are captured by $z(t)$ while fast negative feedback is captured by $y(t)$. Since the parameters are dimensionless, we are free to interpret them in different ways. For example, the parameter $K$ controls the fast negative feedback, and can play the role of a sodium conductance. On the other hand, $\delta$ plays the role of the recovery rate of the slow current. The parameter $x_R$ is the reversal potential of the slow current, and we predict that cardiomyocytes can undergo multiple periodicity changes via a devil's staircase as their potassium reversal potential is shifted towards EAD behavior. These relations could be used to map our findings into the phenomenology of complex models. For example, a devil's staircase-like structure was recently found in complex models of ventricular myocytes[35].

Recent work revealed that shrimps can exhibit quasiperiodic dynamics, characterized by torus-bubbling transitions and multi-tori attractors[3]. This contrasts with the period-doubling structure of periodic shrimps in the Hénon map[4]. In our model, shrimp-shaped regions display a fractal structure with internal isoperiodic stripes forming complete devil's staircases of periodic attractors. Thus, our shrimps are neither isoperiodic nor quasiperiodic, since quasiperiodic attractors are only found at the boundary between steps in the staircase. This expands the previous results, showing that shrimp-shaped regions can either be: (a) isoperiodic[4,5]; or (b) quasiperiodic[3]; or even (c) display infinitely many periodic solutions in the form of a devil's staircase. Moreover, even though the sequence of shrimps in our model form a fractal structure, there is no period-doubling. Instead, the rounded average ISI increases slowly from one shrimp to another without doubling as the shrimps scale down. This pattern is analogous to a harmless devil's staircase in between shrimps.

Along the devil's staircase, we unveiled a qualitative relation between the number of cycles of an attractor (*i.e.*, the maximum period of its corresponding $\{ISI_n\}$ sequence) and the maximum Lyapunov exponent. Quasiperiodic attractors have $L = 0$ and diverging $P$, and they surround each of the steps of the staircase along a given parameter. When this is extended over a bi-parameter region, the quasiperiodic points extend into contour lines for both $-L$ and $P$, generating the striped structure that we observed in our bifurcation diagrams.

EADs can be linked to chaos in complex continuous-time systems[36,37]. In our model, this is not necessary[6]. Although EADs and DADs develop near a chaotic transition into bursting, EADs can be both periodic (existing inside a shrimp) or chaotic. DADs, on the other hand, are chaotic, existing in between shrimps. Other simple map models can be used to simulate cardiac cell behavior at different levels, from the heartbeat interval[38] to the plateau spikes themselves at different levels of complexity[39–41]. Even simple models like the chaotic Rulkov map[42] can exhibit plateau spikes (see, *e.g.*, Fig. 1 in[43]). However, we are not aware of any other map, except ours, that has been thoroughly explored, having its phase diagrams fully traced with the explicit identification of pathological oscillations, like EADs or DADs.

Along the CS-B transition, the membrane potential undergoes a series of infinite periodicity changes before reaching a bursting regime. Some of these changes result in EADs (periodic, quasiperiodic or chaotic) and some in DADs (chaotic). We predict that these transitions are permeated by shrimp-shaped regions, and could also appear in complex cardiac myocyte dynamical models. We could also speculate that chaotic EADs and DADs are more harmful than periodic EADs, although all of them could lead to arrhythmias. Thus, the presence of shrimp-shaped regions with or without devil's staircases could imply a less harmful dynamic. However, these ideas must be thoroughly tested in spatially extended models mimicking the heart tissue. This could have broad applicability and enables experimental validation, enhancing diagnostics, and supporting the development of better tools to treat and prevent cardiac dysfunction.

## DATA AVAILABILITY STATEMENT

Simulations are available in https://github.com/mgirardis/ktz-phasediag

## DECLARATION OF INTERESTS

The authors declare no competing interests.

## ACKNOWLEDGMENTS

M.H.R.T. and M.G.-S. thank financial support from Fundacao de Amparo a Pesquisa e Inovacao do Estado de Santa Catarina (FAPESC), Edital 21/2024 (grant n. 2024TR002507).

## Appendix A: Eckmann-Ruelle method

The Jacobian matrix is defined by the elements $(\mathbf{J}_t)_{ij} = \partial \mathbf{x}_i(t+1)/\partial \mathbf{x}_j(t)$, with $i$ and $j$ equal to 1, 2, or 3, where $\mathbf{x}_i(t)$ is the $i$-th component of $\mathbf{x}$ at time $t$ ($\mathbf{x}_1 = x$; $\mathbf{x}_2 = y$; $\mathbf{x}_3 = z$),

$$\mathbf{J}_t = \begin{bmatrix} \frac{1}{T}F'(u(t)) & -\frac{K}{T}F'(u(t)) & \frac{1}{T}F'(u(t)) \\ 1 & 0 & 0 \\ -\lambda & 0 & 1-\delta \end{bmatrix} , \quad \text{(A1)}$$

with $u(t) = [x(t) - Ky(t) + z(t) + H]/T$ and the derivative $F'(u)$ is

$$F'(u) = \frac{1}{(1+|u|)^2} .$$

The Lyapunov exponents are given by

$$L_i = \lim_{\tau\to\infty} \frac{1}{\tau} \ln|\Lambda_i| , \quad \text{(A2)}$$

where $\Lambda_i$ are the eigenvalues of the product $\mathbf{L}$ of the Jacobian matrices evaluated at each time $t$ from 1 to $\tau$,

$$\mathbf{L} = \mathbf{J}_\tau \mathbf{J}_{\tau-1} \cdots \mathbf{J}_t \cdots \mathbf{J}_2 \mathbf{J}_1 .$$

If the largest exponent, $L = \max_i L_i > 0$, the attractor is said to be chaotic

The calculation in Eq. (A2) is computationally expensive and can be approximated using the Eckmann-Ruelle method[29]. It consists of diagonalizing each Jacobian matrix that makes up $\mathbf{L}$, such that

$$\mathbf{A}_t \mathbf{B}_t = \mathbf{J}_t \mathbf{A}_{t-1} ,$$

$\mathbf{A}_0 = \mathbf{1}$ is the identity matrix, and $\mathbf{A}$ and $\mathbf{B}$ are lower and upper triangular matrices, respectively, obtained by LU decomposition. Thus, we can write $\mathbf{L}$ as

$$\begin{aligned} \mathbf{L} = {} & (\mathbf{A}_\tau \mathbf{B}_\tau \mathbf{A}_{\tau-1}^{-1})(\mathbf{A}_{\tau-1} \mathbf{B}_{\tau-1} \mathbf{A}_{\tau-2}^{-1}) \cdots \\ & \cdots (\mathbf{A}_{t+1} \mathbf{B}_{t+1} \mathbf{A}_t^{-1})(\mathbf{A}_t \mathbf{B}_t \mathbf{A}_{t-1}^{-1}) \cdots \\ & \cdots (\mathbf{A}_3 \mathbf{B}_3 \mathbf{A}_2^{-1})(\mathbf{A}_2 \mathbf{B}_2 \mathbf{A}_1^{-1})(\mathbf{A}_1 \mathbf{B}_1) \\ = {} & \mathbf{A}_\tau \mathbf{B}_\tau \mathbf{B}_{\tau-1} \cdots \mathbf{B}_t \cdots \mathbf{B}_2 \mathbf{B}_1 \\ \approx {} & \mathbf{B}_\tau \mathbf{B}_{\tau-1} \cdots \mathbf{B}_t \cdots \mathbf{B}_2 \mathbf{B}_1 . \end{aligned}$$

The Lyapunov exponents are then approximated by

$$L_i \approx \frac{1}{\tau} \sum_{t=1}^{\tau} \ln|(\mathbf{B}_t)_{ii}| , \ (i = 1, 2, 3 \text{ for } x, y, z) , \quad \text{(A3)}$$

where $\tau$ is a long time (*e.g.*, $\sim 10^7$ ts), and $(\mathbf{B}_t)_{ii}$ are the diagonal elements of the upper triangular matrix $\mathbf{B}_t$.

## Appendix B: Box-count method

The fractal dimension of a set of points can be estimated using the *box-counting method*, a well-established technique in the study of fractal geometry[22,28]. This method involves covering the space containing the data with a series of grids of decreasing box sizes and counting the number of boxes that contain at least one data point.

The core idea is to quantify how the number of occupied boxes $C$ scales with the box size $r$. For a truly fractal structure, this relationship follows a power law:

$$C \sim r^{-D_f} \tag{B1}$$

where $D_f$ represents the fractal dimension. By taking the logarithm of both sides, we obtain a linear relationship in the form

$$\log(C) = -D_f \log(r) + B \ , \tag{B2}$$

where the fractal dimension $D_f$ corresponds to the slope of the line in a log-log plot of $C$ versus $r$.

To implement this method, the $(x_R, w)$ data corresponding to each devil's staircase is first normalized to lie within a unit square, ensuring consistent scaling. Next, a range of box sizes $r$ is selected spanning several orders of magnitude to capture scaling behavior across multiple resolutions. For each box size $r$, a grid is applied, and the number of unique boxes $C$ containing at least one point is counted, yielding the $C \times r$ plot. We fit Eq. (B2) to obtain the slope $D_f$ as the estimate of the fractal dimension. The fit is performed only in the range of $r$ where the relation is linear to avoid boundary effects due to resolutions that are either too fine or too coarse. Logarithmic spacing of $r$ is often preferred to ensure even coverage across different orders of magnitude.

## Appendix C: Correlation coefficient

We used MATLAB® function *corrcoef(A,B)* to calculate the correlation coefficient and the significance. The method works as follows. The correlation coefficient $R$ of two curves is a measure of their linear dependence. If each curve has $M$ points, then the Pearson correlation coefficient between curves $A$ and $B$ is defined as

$$R = \frac{1}{M-1}\sum_{k=1}^{M}\left(\frac{A_k-\mu_A}{\sigma_A}\right)\left(\frac{B_k-\mu_B}{\sigma_B}\right) \ , \tag{C1}$$

where $A_k$ and $B_k$ are the points belonging to each curve, and $\mu_X$, $\sigma_X$ are, respectively, the mean and standard deviation of each of the curves, $X = A$ or $X = B$. For example, we use $A = -L$ as the negative of the Lyapunov exponent, and $B = 1/P$ as the inverse of the number of cycles of the periodic attractor. Both curves are evaluated along the $x_R$ axis, having all the other parameters fixed.

The significance of the correlation is estimated by calculating the $t$-statistic with $M-2$ degrees of freedom,

$$t = \frac{R\sqrt{M-2}}{1-R^2} \ . \tag{C2}$$

The probability $p$ that two random curves $A$ and $B$ present a $t$ given by Eq. (C2), is then calculated by

$$p = 2\min[\mathcal{P}(t; M-2); 1-\mathcal{P}(t; M-2)] \ ,$$

where $\mathcal{P}(t;\nu)$ is the cumulative Student's $t$-distribution with $\nu$ degrees of freedom[44]. In other words, $\mathcal{P}(t;\nu)$ is the probability of finding some $s < t$,

$$\mathcal{P}(t;\nu) = \frac{1}{\sqrt{\nu\pi}}\frac{\Gamma\left(\frac{\nu+1}{2}\right)}{\Gamma\left(\frac{\nu}{2}\right)}\int_{-\infty}^{t}\left(1+\frac{s^2}{\nu}\right)^{-\frac{\nu+1}{2}} ds \ ,$$

where $\Gamma(\cdot)$ is the Gamma function. Therefore, $p$ is the probability of being in one of both tails of the Student's $t$-distribution. This essentially amounts to a hypothesis testing, telling us how likely it is ($p$) for two random curves to present a correlation coefficient $R$. The smaller the calculated $p$, the less likely $R$ was found by chance, and hence the more significant is the correlation $R$ between $A$ and $B$.